\DeclareUnicodeCharacter{2212}{\textendash}
\documentclass[prd,twocolumn,floatfix,amsmath,nofootinbib,amssymb,floatfix]{revtex4}
\usepackage{graphicx,color,dcolumn,booktabs,bm}
\definecolor{plum}{RGB}{221,160,221} 
\usepackage{longtable,lscape}
\usepackage{natbib}
\usepackage{txfonts}
\usepackage{overpic}
\usepackage{amssymb}
\usepackage{indentfirst}
\usepackage{xcolor}
\usepackage{cases}
\usepackage{color}
\usepackage{multirow}
\usepackage{makecell}
\usepackage{enumerate}
\usepackage{adjustbox}
\usepackage{graphicx,color,dcolumn,booktabs,bm}
\usepackage{epsfig}
\usepackage[colorlinks, citecolor=blue,anchorcolor=red,menucolor=red, linkcolor=red,filecolor=red,urlcolor=blue,frenchlinks=red]{hyperref}
\newcommand{\hd}{{\overline D}}
\newcommand{\f}{\frac}
 
\newcommand{\smallw}{{\scriptscriptstyle W}}

\newcommand{\muw}{\mu_\smallw}

\newcommand{\mub}{\mu_b}

\newcommand{\dnp}{\delta^{{\scriptscriptstyle NP}}}

\newcommand{\vt}{v_\Delta}

\newcommand{\beq}{\begin{eqnarray}}
	\newcommand{\eeq}{\end{eqnarray}}

\def\sm{\textit{$\mathcal{SM}\;$}}
\def\bsm{\textit{$\mathcal{BSM}\;$}}
\graphicspath{{Figures/}} %

\begin{document}
\title{ The two Higgs doublet type-II seesaw model: Naturalness and $\bar{B}\to X_s\gamma$ versus heavy Higgs masses}
\author{B.~Ait Ouazghour$^{1}$}\email{brahim.aitouazghour@edu.uca.ac.ma}
\author{M.~Chabab$^{1}$}\email{ mchabab@uca.ac.ma  (Corresponding author)}
\affiliation{
$^1$LPHEA, Faculty of Science Semlalia, Cadi Ayyad University, P.O.B. 2390 Marrakech, Morocco.
}

\begin{abstract}
We extend the work \cite{Ouazghour:2018mld} to a more general and detailed analysis through studying the naturalness problem and $B$ physics constraints within the context of two higgs doublets model augmented with a  complex  scalar triplet field ($2HDMcT$). We first derive the modified Vetman conditions and show that naturalness problem might be evaded  at the electroweak scale. Then, we evaluate the branching ratio of $\bar{B}\to X_s\gamma$ radiative decay at the Next to leading order. Besides unitarity, boundedness from below constraints and the present collider data from Higgs signal rate measurements, the parameter space of $2HDMcT$ is then revisited to see how the modified Veltman conditions and $\bar{B}\to X_s\gamma$ experimental bounds induce significant delimitations. Our analysis shows that the naturalness affects drastically the masses of heavy Higgs $h_3$, $A_2$, $H_2^\pm$ and $H^{\pm\pm}$. More specifically, we considered three benchmak scenarios corresponding to $\mu_2 > 20, 40, 70$ GeV. For $\mu_2 > 40$, we find stringent upper limits  on $m_{h_2}$, $m_{A_1}$  and $m_{H_1^\pm}$ at $880$, $920$ and $918$ GeV respectively. If, in addition, the measurements of $\bar{B}\to X_s\gamma$ decay rate are also considered, the lower masses limits of nonstandard  Higgs  bosons ($h_2$, $h_3$, $A_1$, $A_2$, $H_1^\pm$, $H_2^\pm$, $H^{\pm\pm}$) are pushed up to higher values located between $490$ and $750$ GeV. Besides, we also demonstrate that the $\bar{B}\to X_s\gamma$ experimental limits can only be accommodated within $2HDMcT$ if two conditions are fullfilled : $\alpha_1$ parameter positive  and the Higgs mass hierarchy $m_{h_{3}}>m_{H_2^\pm}>m_{H^{\pm\pm}}$. 
\end{abstract}
\maketitle
\section{ Introduction }
\label{sec:intro}
\paragraph*{}
Since the discovery by the ATLAS and CMS experiments at the LHC \cite{ATLAS:2012yve,aad2013atlas,CMS:2012qbp,chatrchyan2013observation} of a particle with properties that are remarkably consistent with the Higgs boson of the Standard Model  (\sm) \cite{gunion2018higgs} with a mass about $125$ GeV \cite{ATLAS:2015yey}, ongoing experimental tests of the \sm theory and its limitations have provided stringent constraints on new physics beyond the Standard Model (\bsm). Indeed, despite its brilliant success to describe fundamental particle and their interactions, many questions still remain unexplained and/or controversial., e.g. the dark matter existence in the universe, neutrino oscillations and mass hierarchy. Amidst all of that, the naturalness problem \cite{Pivovarov:2007dj} is also far from being resolved.

It is then legitimate to propose other theoretical perspectives which could confirm the pattern expected from the Brout-Englert-Higgs mechanism and provide a convincing interpretation of new phenomena \bsm. For the electroweak hierarchy problem, given the absence of supersymmetry at $TeV$ scale, a plethora of novel altenatives have been considered so far, more particularly with an extended Higgs sector. It is known that one has to call upon new physics where the new degrees of freedom in a particular \bsm model conspire with those of the Standard Model to get rid of the quadratic divergencies modifying the Veltman condition $(VC)$. As examples of such \bsm approaches, we can cite the \sm  extended with singlet  or triplet scalar fields \cite{Chabab:2015nel, Chabab:2018ert, Grzadkowski:2009bp, Karahan:2014ola, Das:2023tna}, the two Higgs doublet \cite{Darvishi:2017bhf}, and the recent analysis within  $2HDM$ augmented with a real triplet scalar \cite{Ait-Ouazghour:2020slc}. 
\paragraph*{}
Due to the similarity in the mass generation mechanism between type-II seesaw and the Brout-Englert-Higgs mechanism,  the two Higgs double model extension to the type-II seesaw model ($2HDcT$) and its collider phenomenology are quite appealing, displaying some phenomenological characteristics especially different from those emerging  in $2HDM$ scalar sector.  Apart its broader spectrum than $2HDM$'s one, the doubly charged Higgs the $H^{++}$ is the smoking gun of $2HDcT$.  The search for $H^{++}$ is intensively undergoing by ATLAS and CMS, via the decay channels to $4$ leptons. Searches for $H^{++}H^{--}$ and $H_i^{+}H_i^{-}$,  with $i=1,2$, decaying to the same sign di-lepton, are among the most promising discovery channels of this model. Another prospective signal is $pp \to Z/\gamma \to H_2^+ H_2^- \to H^{++} W^- H^{--} W^+ \to l^+l^+l^-l^-+4j$. This process does not show up neither in $2HDM$ nor  in the Higgs triplet model ($HTM$). To shed  light on the $2HDMcT$ model's significance with respect to both the  $2HDM$ and  $HTM$,  we have conducted in \cite{Ouazghour:2018mld} a first study of  the production and decays of  several Higgs processes. Our  calculations of the cross sections and branching ratios, along with the comprehensive analyses of various decay channels have carried out in-depth exploration of the intricate decay modes of $h_{2,3}$ and $A_{1,2}$, with a specific emphasis on the final states $\gamma\gamma$ and $\tau\tau$. The results from the diphoton decay channels of the pseudoscalar Higgs ($A_1, A_2$), revealed significant suppressions in the products of effective cross sections and branching ratios compared to experimental limits. Furtheremore, we have also investigated scenarios involving the production of Higgs pairs, with a focus on the cases where the observed Higgs boson corresponds either to the lighter scalar ($h_1$) or to the second lightest scalar ($h_2$). On the other hand, besides Higgs phenomenology,  interactions between doublet and triplet fields may also induce a strong first order electroweak phase transtion, thus providing conditions for the baryon asymmetry generation via electroweak baryongenesis (see  \cite{Ramsey-Musolf:2019lsf} for more details).

\paragraph*{}
In this paper, we aim to investigate both the naturalness problem and $\bar{B}\to X_s\gamma$ decay rate in the context of a two Higgs doublet type-II seesaw model, dubbed $2HDMcT$.  Besides explaining neutrino oscillations, this model  has also been examined either to deal with the dark matter issue \cite{Chen:2014lla}, or to perform phenomenological analyses \cite{Ouazghour:2018mld, Chen:2014xva}. More precisely, we will first derive the Veltman conditions ($VC$), study how to soften their divergencies to gain insight into $2HDMcT$ parameter space as on the allowed masses of the heavy scalars in the Higgs sector. Next, we will evaluate the $\bar{B}\to X_s\gamma$ decay rate at the next to leading order $(NLO)$ and see whether constraints from the $\bar{B}\to X_s\gamma$ measurements affect the $2HDMcT$ spectra.  Still, it remains to check that the implementation  for$mVC$ and $\bar{B}\to X_s\gamma$ constraints are consistent with theoretical requirements and experimental data and how they reshape the Higgs spectrum of our model, in comparison with the previous phenomenological analysis reported in \cite{Ouazghour:2018mld}

\paragraph*{}
This work is organized as follows: In Sec. \ref{prese_2HDMcT}, we  briefly review the main features of two Higgs Doublet Type-II Seesaw model. In Sec. \ref{con_the_exp}, we briefly present the theoretical  and experimental constraints. Section \ref{VC} is devoted to the derivation of the modified Veltman condition ($VC$) in the $2HDMcT$. In Sec. \ref{BtoXga}, we discuss the constraints on the parameter space from $\bar{B}\to X_s \gamma$. The analysis and discussion of the results are performed in Sec. \ref{result}, with emphasis on the effects of the modified Veltman conditions as well as the constraints from  the $\bar{B}\to X_s\gamma$ on the heavy Higgs spectrum, essentially  on the charged Higgs sector.  A summary of our results will be drawn in Sec. \ref{conlusion}.

\section{TYPE II SEESAW MODEL: BRIEF REVIEW}
\label{prese_2HDMcT}
	The $2HDMcT$ contains two Higgs doublets $H_{i}$ (i = 1,2) 	and one colorless scalar field $\Delta$ transforming as a triplet under the $SU(2)_L$ gauge group with hypercharge $Y_\Delta=2$. In this case, the most general gauge-invariant Lagrangian of $2HDMcT$ is given by, 
%
\small{\begin{equation}
		\begin{matrix}
			\mathcal{L}=\sum_{i=1}^2(D_\mu{H_i})^\dagger(D^\mu{H_i})+Tr(D_\mu{\Delta})^\dagger(D^\mu{\Delta})\vspace*{0.12cm}\\
			\hspace{-3cm}-V(H_i, \Delta)+\mathcal{L}_{\rm Yukawa}
			\label{eq:thdmt-lag}
		\end{matrix}
\end{equation}}
where the covariant derivatives are defined as,
\begin{equation}
	D_\mu{H_i}=\partial_\mu{H_i}+igT^a{W}^a_\mu{H_i}+i\frac{g'}{2}B_\mu{H_i} \label{eq:covd1}
\end{equation}
\vspace*{-0.6cm}
\begin{equation}
	D_\mu{\Delta}=\partial_\mu{\Delta}+ig[T^a{W}^a_\mu,\Delta]+ig' \frac{Y_\Delta}{2} B_\mu{\Delta} \label{eq:covd2}
\end{equation}
(${W}^a_\mu$, $g$), and ($B_\mu$, $g'$) denoting respectively the $SU(2)_L$ and $U(1)_Y$ gauge fields and couplings and $T^a \equiv \sigma^a/2$, with $\sigma^a$ ($a=1, 2, 3$)  the Pauli matrices. In terms of the two $SU(2)_L$ Higgs doublets $H_i$ and the triplet field $\Delta$, the $2HDMcT$ scalar potential is given by \cite{Ouazghour:2018mld, Chen:2014xva}:
\begin{widetext}
	\[
	\begin{matrix}
		V(H_i,\Delta)&=&m^2_{1}\, H_1^\dagger H_1 + m^2_{2}\, H_2^\dagger H_2- m_{3}^2\,H_1^\dagger H_2\hspace*{0cm}-\hspace*{0cm}m_{12}^2\,H_2^\dagger H_1 +\frac{\lambda_1}{2} (H_1^\dagger H_1)^2 + \frac{\lambda_2}{2}  (H_2^\dagger H_2)^2\hspace*{0cm}+\hspace*{0cm}\lambda_3\, H_1^\dagger H_1\, H_2^\dagger H_2\nonumber\\
		&& + \lambda_4\, H_1^\dagger H_2\, H_2^\dagger H_1+\frac{\lambda_5}{2}\hspace*{0cm}*\hspace*{0cm} \left[(H_1^\dagger H_2)^2+ (H_2^\dagger H_1)^2 \right]+\hspace*{0cm}\lambda_6\,H_1^\dagger H_1 Tr\Delta^{\dagger}{\Delta} +\lambda_7\,H_2^\dagger H_2 Tr\Delta^{\dagger}{\Delta}\hspace*{0cm}\hspace*{1.2cm}\nonumber\\
		&&+[\mu_1 H_1^T{i}\sigma^2\Delta^{\dagger}H_1 + \mu_2H_2^T{i}\sigma^2\Delta^{\dagger}H_2 + \mu_3 H_1^T{i}\sigma^2\Delta^{\dagger}H_2  + {\rm h.c.}]\hspace*{0cm}+\hspace*{0cm}\lambda_8\,H_1^\dagger{\Delta}\Delta^{\dagger} H_1\hspace*{2.5cm}\nonumber\\
		&& + \lambda_9\,H_2^\dagger{\Delta}\Delta^{\dagger} H_2+\hspace*{0cm}m^2_{\Delta}\, Tr(\Delta^{\dagger}{\Delta}) +\bar{\lambda}_8(Tr\Delta^{\dagger}{\Delta})^2\hspace*{0cm}+\hspace*{0cm}\bar{\lambda}_9Tr(\Delta^{\dagger}{\Delta})^2\hspace*{4.6cm}
		\label{eq:VDelta}
	\end{matrix}
	\]
\end{widetext}
where $Tr$ denotes the trace over 2x2 matrices. The triplet $\Delta$ and doublet Higgs H are represented by, 
\begin{eqnarray}
	\Delta &=&\left(
	\begin{array}{cc}
		\delta^+/\sqrt{2} & \delta^{++} \\
		(v_t+\delta^0+i\eta_0)/\sqrt{2} & -\delta^+/\sqrt{2}\\
	\end{array}
	\right)\end{eqnarray}
\vspace*{-0.25cm}
\begin{eqnarray}
	H_1&=&\left(
	\begin{array}{c}
		\phi_1^+ \\
		\phi^0_1 \\
	\end{array}
	\right){,}~~~H_2=\left(
	\begin{array}{c}
		\phi_2^+ \\
		\phi^0_2 \\
	\end{array}
	\right)\end{eqnarray}
with $\phi^0_1=(v_1+\rho_1+i\eta_1)/\sqrt{2}$ and $\phi^0_2=(v_2+\rho_2+i\eta_2)/\sqrt{2}$. After the spontaneous electroweak symmetry breaking, the
Higgs doublets and triplet fields acquire their vacuum expectation values, respectively dubbed $v_1$, $v_2$ and $v_t$, and eleven physical Higgs
states appear, namely: three CP-even neutral Higgs bosons $(H_1, H_2, H_3)$, four simply charged Higgs bosons $(H_1^{\pm}, H_2^{\pm})$,  two CP odd Higgs $(A_1, A_2)$, and finally two doubly charged Higgs bosons $H^{\pm\pm}$. For detail see \cite{Ouazghour:2018mld}.

\section{THEORETICAL AND EXPERIMENTAL CONSTRAINTS}
\label{con_the_exp}
\paragraph*{}
The allowed $2HDMcT$ parameter space must generally obey to many theoretical and experimental constraints. These  constraints include vacuum stability, unitarity  and perturbativity, in addition to  experimental constraints and exclusion limits especially originating from  the measurements of Higgs boson properties, flavor-changing neutral currents and electroweak precision observables. These are  smoking guns to probe imprints of new physics within the model. Hereafter, we provide a brief description of all constraints invoked to delimit the model parameter space and used in our subsequent analysis \cite{Ouazghour:2018mld}:
	\begin{itemize}
	\item \textbf{Unitarity:} The scattering processes have to be unitary.
	\item \textbf{Perturbativity:} The quartic couplings of scalar potential must obey the following conditions:$| \lambda_i|<8 \pi$ for each $i=1,..,9$ and $| \bar{\lambda}_i|<8 \pi$ for $i=8,9$.
	\item \textbf{Vacuum stability} : Positivity in any direction of fields $\Phi_i$, $\Delta$ and boundedness from below $BFB$.
	\item {\bf Experimental Higgs boson exclusion limits:} 
	Existing exclusion limits at the 95\% confidence level from Higgs
	searches at LEP, LHC, and Tevatron have been enforced via  \texttt{HiggsBounds-5.10.1} \cite{bechtle2010higgsbounds,bechtle2011higgsbounds,bechtle2014higgsbounds,bechtle2020higgsbounds,bahl2022testing}.
	\item {\bf SM-like Higgs boson discovery:} Compatibility with the measurements of the Higgs signal rate
	 from various searches at the $95 \%$ confidence level, essentially from the
	LHC Run 2. Here we used  \texttt{HiggsSignals-2.6.1} \cite{bechtle2014higgssignals,bechtle2021higgssignals}.	
	\item {\bf The electroweak precision observables $(EWPO)$:} Generally, the additional scalars in a model introduce extra correctionsto the gauge boson self energy diagrams, which affect electroweak precision observables $(EWPO)$, as the oblique parameters $S$, $T$ and $U$. Accordingly the new charged, doubly charged and neutral bosons in $2HDMcT$ yield corrections to the analytic formulas of $S$  and $T$. To estimate the oblique parameters in the present model, we use the prescription adopted in \cite{Lavoura:1993nq}  where for simplicity  the general expressions of $S$, $T$ and $U$ where established  by assuming  that the complex scalar multiplet with a small vacuum expectaion value and no coupling to the other scalars of the theory. This translates in $2HDMcT$ to $v_t \le 1$ GeV with the following rotation matrix elements: $\alpha_2$ and $\alpha_3$ almost vanishing, while  $C_{33}$ (charged) and $O_{33}$ tend to $\approx 1$.
	 In this context, the doublets  and triplet fields decouple. This means that major contributions to the physical fields $h_3$, $A_2$, $H^\pm_2$ and $H^{\pm\pm}$ originate from triplet fields while the dominat contributions to  $h_1$, $h_2$, $A_1$ and $H_1^\pm$  are from doublet fields. Using the general expressions presented in \cite{Grimus:2007if,Grimus:2008nb, Lavoura:1993nq}, we readily  calculate the new contribution to the $S$ and $T$ parameters from the new scalars in $2HDMcT$  as,
	
	\begin{eqnarray} 
		\hspace*{-10cm}T&=&\frac{1}{16 \pi m_W^2 s_W^2}(F(m_{H^{\pm\pm}}^2,m_{H^\pm_2}^2)+F(m_{H^\pm_2}^2,m_{h_3}^2)+{\cal R}_{21}^2 F(m_{H_1^\pm}^2,m_{h_1}^2) \nonumber\\
		&+&{\cal R}_{22}^2 F(m_{H_1^\pm}^2,m_{h_2}^2)+F(m_{H_1^\pm}^2,m_{A_1}^2)-{\cal R}_{21}^2 F(m_{h_1}^2,m_{A_1}^2)-{\cal R}_{22}^2 F(m_{h_2}^2,m_{A_1}^2)\nonumber\\
		& +&3({\cal R}_{11}^2 (F(m_Z^2,m_{h_1}^2)-F(m_W^2,m_{h_1}^2))+{\cal R}_{12}^2 (F(m_Z^2,m_{h_2}^2)-F(m_W^2,m_{h_2}^2)))   \nonumber \\
		&-& 3(F(m_Z^2,m_{h_{ref}}^2)-F(m_W^2,m_{h_{ref}}^2))) 
	\end{eqnarray} 
	and
	\begin{align}
		\hspace*{-10cm}S= &\frac{1}{24 \pi}[(2s_W^2-1)^2G(m_{H_1^+}^2,m_{H_1^+}^2,m_Z^2)+{\cal R}_{21}^2 G(m_{h_1}^2,m_{A_1}^2,m_Z^2)\nonumber  \\
		&+({\cal R}_{11}^2+{\cal R}_{21}^2)\ln(m_{h_1}^2)+({\cal R}_{12}^2+{\cal R}_{22}^2)\ln(m_{h_2}^2)+\ln(m_{A_1}^2))  \nonumber \\ 
		&+{\cal R}_{22}^2 G(m_{h_2}^2,m_{A_1}^2,m_Z^2)-2 \ln(m_{H_1^{\pm}}^2)-\ln(m_{h_{ref}}^2)\nonumber \\ 
		&+{\cal R}_{11}^2 \hat G(m_{h_{1}^2},m_Z^2)+{\cal R}_{12}^2 \hat G(m_{h_{2}}^2,m_Z^2)-\hat G(m_{h_{ref}}^2,m_Z^2)]+\\
		&\frac{1}{3\pi}\bigg[-\ln\left(\frac{m_{H^{\pm\pm}}^2}{m_{h_3}^2}\right)+\frac{1}{2}\xi\left(\frac{m_{h_3}^2}{m_Z^2}\right)+\frac{(1-2s^2_w)^2}{2}\xi\left(\frac{m_{H^{\pm\pm}}^2}{m_Z^2}\right)\nonumber\\
		&+\frac{s^4_w}{2} \xi\left(\frac{m_{H_2^{\pm}}^2}{m_Z^2}\right)\bigg]	
	\end{align} 
	
	with
	\begin{equation}
		{\cal R}_{11} = C_{11}R_{11}+C_{12}R_{12}
	\end{equation} 
	\begin{equation}
		{\cal R}_{22} = C_{21}R_{21}+C_{22}R_{22}
	\end{equation} 
	\begin{equation}
		{\cal R}_{12} = C_{11}R_{21}+C_{12}R_{22}
	\end{equation} 
	\begin{equation}
		{\cal R}_{21} = C_{21}R_{11}+C_{22}R_{12}
	\end{equation} 
	and $m_{h_{ref}}$ is the reference mass of the neutral SM Higgs and $s_w = sin(\theta_w)$ where $\theta_w$ is the Weinberg angle. The explicit forms of these functions, $F \left( x, y \right)$, $G \left( I, J, Q \right)$, $\hat G \left( I, Q \right)$ and $\xi(x)$
	are given in \cite{grimus801oblique,Lavoura:1993nq}.
	\begin{equation}
		F \left( x, y \right) \equiv
		\left\{ \begin{array}{ll}
			{\displaystyle \frac{x+y}{2} - \frac{xy}{x-y}\, \ln{\frac{x}{y}}}
			&\Leftarrow\ x \neq y,
			\\*[3mm]
			0 &\Leftarrow\ x = y.
		\end{array} \right. 
		\label{F}
	\end{equation}
	\begin{align}
		G \left( I, J, Q \right) = &- \frac{16}{3} + \frac{5 \left( I + J \right)}{Q} - \frac{2 \left( I - J \right)^2}{Q^2}\nonumber\\ 
		&	+\frac{3}{Q}\left[ \frac{I^2 + J^2}{I - J}-\frac{I^2 - J^2}{Q}+\frac{\left( I - J \right)^3}{3 Q^2} \right]
		\ln{\frac{I}{J}}+\frac{r}{Q^2}\, f \left( t, r \right).
		\label{G1}
	\end{align} 
	If  $I=J$, $G(I,J,Q)$ is :
	\begin{eqnarray}
		G \left( I, J, Q \right) &=& - \frac{16}{3} +\frac{16}{Q} I
		+ \frac{r}{Q^3}\, f \left( t, r \right) \nonumber 
	\end{eqnarray}
	and :
	\begin{eqnarray}
		\label{f}
		f \left( t, r \right) \equiv \left\{ \begin{array}{lcl}
			{\displaystyle
				\sqrt{r}\, \ln{\left| \frac{t - \sqrt{r}}{t + \sqrt{r}} \right|}
			} & \Leftarrow & r > 0,
			\\*[3mm]
			0 & \Leftarrow & r = 0,
			\\*[2mm]
			{\displaystyle
				2\, \sqrt{-r}\, \arctan{\frac{\sqrt{-r}}{t}}
			} & \Leftarrow & r < 0.
		\end{array} \right. \nonumber
	\end{eqnarray}
	with :
	\begin{eqnarray}
		\label{tr}
		t \equiv I + J - Q
		\quad \mbox{and} \quad
		r \equiv Q^2 - 2 Q \left( I + J \right) + \left( I - J \right)^2 \nonumber
	\end{eqnarray}
	\begin{eqnarray}
		\hat G \left( I, Q \right) &=&- \frac{79}{3} + 9\, \frac{I}{Q} - 2\, \frac{I^2}{Q^2}+\bigg( 12 - 4\, \frac{I}{Q} + \frac{I^2}{Q^2} \bigg)
		\frac{f \left( I, I^2 - 4 I Q \right)}{Q}\nonumber \\ 
		&+& \bigg( - 10 + 18\, \frac{I}{Q} - 6\, \frac{I^2}{Q^2} + \frac{I^3}{Q^3}
		- 9\, \frac{I + Q}{I - Q} \bigg) \ln{\frac{I}{Q}}. 
		\label{G2}
	\end{eqnarray}
	and
	\begin{equation}
		\begin{aligned}
			\xi(x) &= 12\left(\frac49 - \frac43x + \frac{1}{12}(4x-1)f(x)\right),\\
			f(x) &= \left\{\begin{aligned}&-4\sqrt{4x-1}\arctan \frac{1}{\sqrt{4x-1}},\quad \text{for }4x>1,\\
				&\sqrt{1-4x}\ln\frac{2x-1+\sqrt{1-4x}}{2x-1-\sqrt{1-4x}},\quad \text{for }4x\le1,\end{aligned}\right.
		\end{aligned}
		\label{X}
	\end{equation}
	By assuming that $U=0$,  analysis of the precision electroweak data with the new PDG mass of the $W$ boson yields  \cite{ParticleDataGroup:2020ssz}:
	\begin{equation}
		S^{exp} = 0.05 \pm 0.08,\;\;\;T^{exp} = 0.09 \pm 0.07\;\;\; and \;\;\; \rho_{ST} = +0.92.
	\end{equation} 
	where $\rho_{ST}$ corresponds to the correlation coefficient in the $\chi_{ST}^2$ analysis. Next, we use the following $\chi^2_{ST}$ test where only points that are within $95\%$ confidence level (C.L.) of the PDG measurements are considered,\\

\begin{eqnarray}
\chi^2_{ST} &=& \frac{1}{\hat{\sigma_T}^2_{1}(1-\rho_{ST}^2)}(T - T^{exp})^2
+ \frac{1}{\hat{\sigma_S}^2_{1}(1-\rho_{ST}^2)}(S - S^{exp})^2 \nonumber \\ 
& -& \frac{2\rho_{ST}}{\hat{\sigma}_{T}\hat{\sigma}_{S}(1-\rho_{ST}^2)}(T - T^{exp})(S - S^{exp})\le R^2\,, \label{eq:chi2}
\end{eqnarray}
with $R^2=2.3$, $5.99$ and $11.83$ at $68.3 \%$,
$95 \%$ and $99.7 \%$  CLs, respectively and $\hat{\sigma}_{T}$, $\hat{\sigma}_{S}$ are one-sigma errors.	
\end{itemize}

\section{THE MODIFIED VELTMAN CONDITIONS}
\label{VC}
\paragraph*{}
As it is known, the new degrees of freedom in any new physics models spectrum conspire with the $SM$ ones in order to soften the quadratic divergencies. The Veltman conditions (VC) are then modified and naturalness problem gets under control within the model parameter space.  So, to derive the Veltman conditions in $2HDMcT$, one just has to collect the quadratic divergencies of the Higgs self-energies in terms of the original fields, namely the doublet $\Phi_1$ , $\Phi_2$ and triplet $\Delta$, without spontaneous breaking of the $SU(2)\times U(1)$ gauge symmetry \cite{Veltman:1980mj}. There are various ways to do that, however the safest way is to use  the dimensional regularization \cite{siegel1979supersymmetric,einhorn1992effective,al1992quadratic} as this approach guarantees both gauge Lorentz invariances. Besides, we performed this calculation in a general linear $R_\xi$ gauge and verified that the results obtained are entirely free of $\xi$-parameter as required. To work out these quadratic divergencies, we followed exactly the calculation steps described in our previous work on the two Higgs doublet model extended by a real triplet scalar field $\Delta$ \cite{Ait-Ouazghour:2020slc} and in  \cite{Newton:1993xc,newton1994can, Al-Sarhi:1990nmv, Einhorn:1992um}. Throughout these calculations, we used the following convenient notation,
\begin{itemize}
	\item[$\bullet$] $(\Phi)_p$ and $(\Delta)_q$ denotes the p-component of the doublets and q-component of the triplet fields.
	\item[$\bullet$] $I_{11}(I_{22})$ as the quadratically divergent part of the two-point functions with either the upper or lower components of $\Phi_1$ ($\Phi_2$) fields, on both external lines of the relevant Feynman Diagrams. Similarly, we label $I_{33}$ for the triplet two-point functions, with one of $\Delta$ components on external lines.
\end{itemize}
To derive the final results in symmetry unbroken phase, it is necessary to summarize all possible schemes, taking only the coefficients of divergent parts in $I_{ij}$, leading to:
\begin{figure*}[hbtp]
	\centering
	\resizebox{0.24\textwidth}{!}{
			\includegraphics{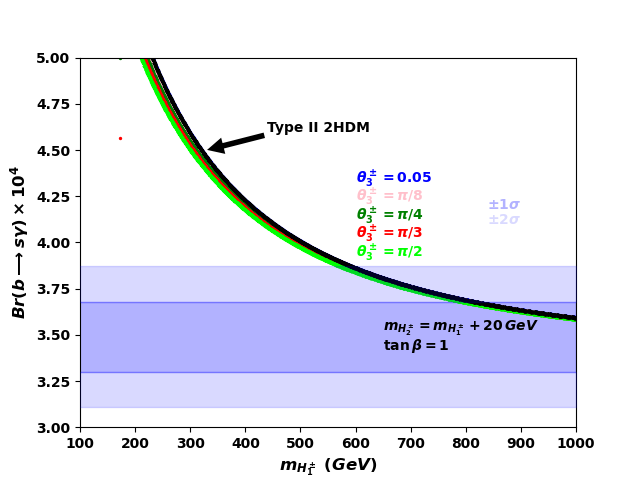}}
	\resizebox{0.24\textwidth}{!}{
	\includegraphics{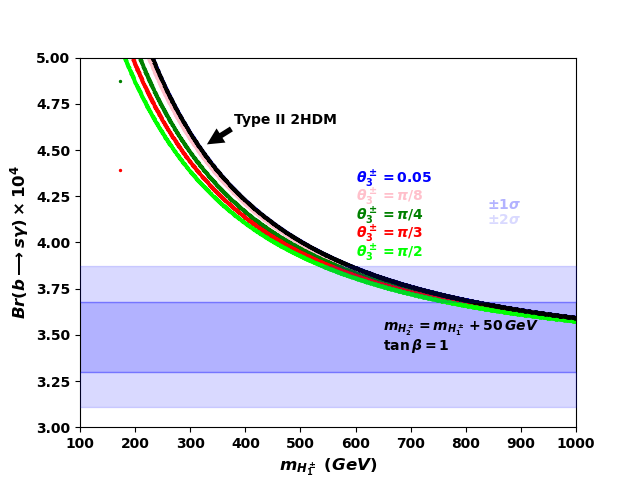}}
\resizebox{0.24\textwidth}{!}{
\includegraphics{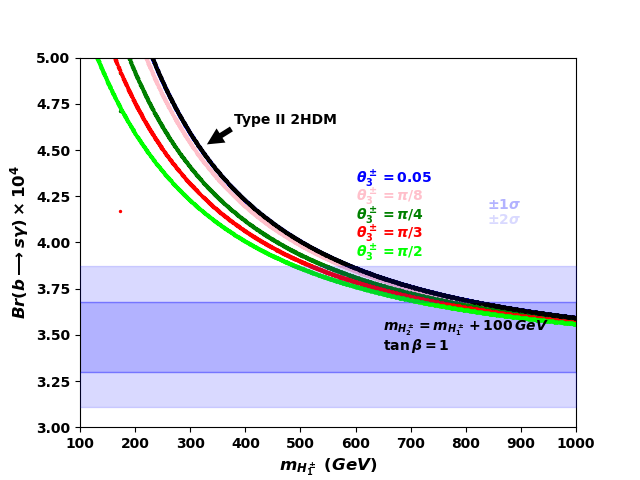}}
\resizebox{0.24\textwidth}{!}{
\includegraphics{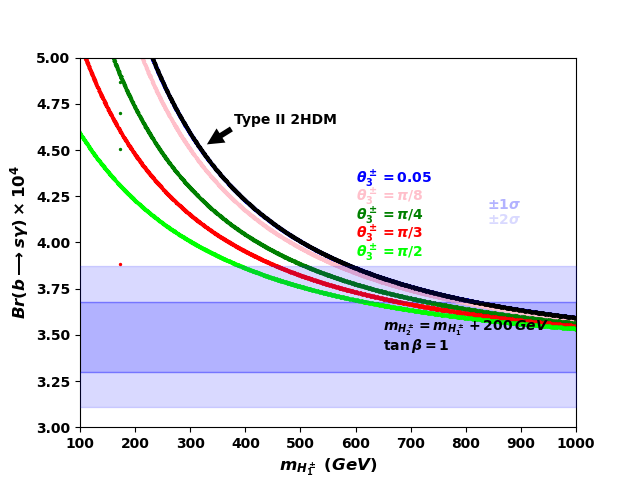}}

\resizebox{0.24\textwidth}{!}{
\includegraphics{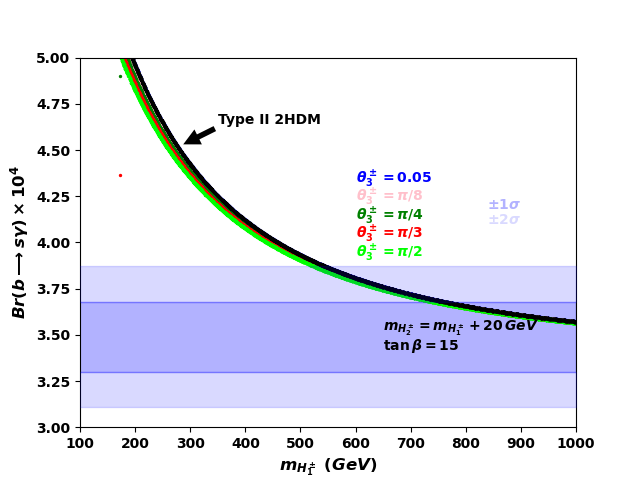}}
\resizebox{0.24\textwidth}{!}{
\includegraphics{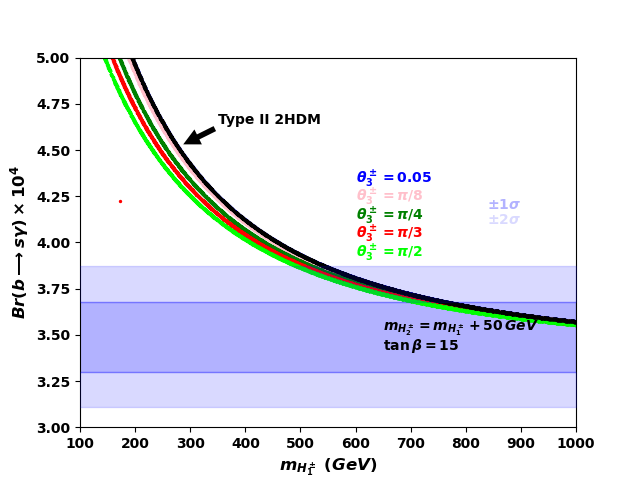}}
\resizebox{0.24\textwidth}{!}{
\includegraphics{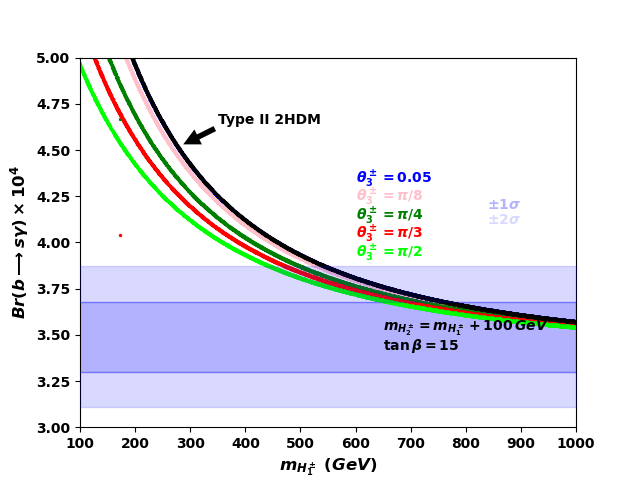}}
\resizebox{0.24\textwidth}{!}{
\includegraphics{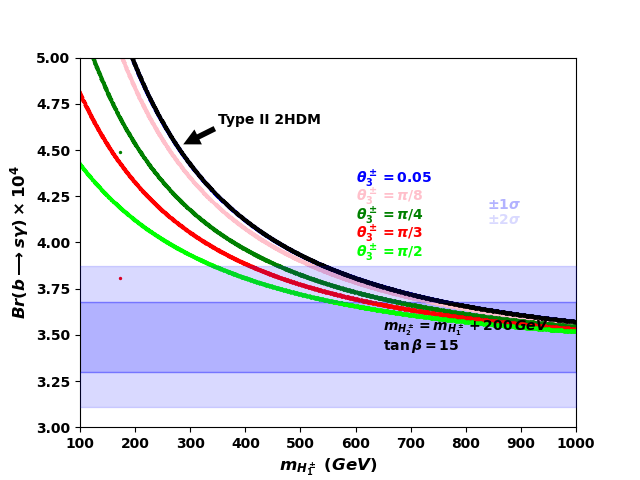}}
	
	\caption{Predictions for the branching ratio of $B \rightarrow X_s \gamma$ in $2HDMcT$ type II as a function of $m_{H_1^\pm}$. First and second column panels correspond respectively to $\tan \beta=1\;and\;\;15$, with $m_{H_2^\pm}-m_{H_1^\pm}=20,\;50,\;100$ and $200$ GeV.}
	\label{fig:constraints}
	
\end{figure*}
\begin{eqnarray}
	I_{11}&\rightarrow&3/4g^2+1/4g^{'2}+3\lambda_1+2\lambda_3+\lambda_4\nonumber\\&&+(3\lambda_6+\frac{3}{2}\lambda_8)-\left(\frac{\sqrt{2}}{v}\right)^2 \frac{m_D^2}{cos^2\beta}
\end{eqnarray}
\begin{eqnarray}
	I_{22}&\rightarrow&3/4g^2+1/4g^{'2}+3\lambda_2+2\lambda_3+\lambda_4\nonumber\\&&+(3\lambda_7+\frac{3}{2}\lambda_9)-\left(\frac{\sqrt{2}}{v}\right)^2 \frac{m_U^2}{sin^2\beta}
\end{eqnarray}
\begin{eqnarray}
	I_{33}&\rightarrow&2g^2+g^{'2}+2(\lambda_6+\lambda_7)+\lambda_8+\lambda_9+8\bar{\lambda}_8+6\bar{\lambda}_9
\end{eqnarray}
where $m^2_D=m^2_e+m^2_\mu+m^2_\tau+3(m^2_d+m^2_s+m^2_b)$, $m^2_U=3(m^2_u+m^2_c+m^2_t)$ and $v^2=v_1^2+v_2^2+2v_t^2$ is the
square of the Standard Model vacuum expectation value $(\approx246 GeV)$.

\paragraph*{}
Finally, thanks to the relations $\cos \beta=\frac{v_1}{v_0}$, $\sin \beta=\frac{v_2}{v_0}$, $m_W=\frac{gv}{2}$, $e=g \sin\theta_{weinberg}$ and $g^{'}=g \tan \theta_{weinberg}$, the tadpoles equations in the broken phase are thoroughly recovered:
\begin{eqnarray}
\delta T_{d_1}&=&\frac{m_W^2}{v^2}\left(2+\frac{1}{c_w^2}\right)+3\lambda_1+2\lambda_3+\lambda_4+(3\lambda_6+\frac{3}{2}\lambda_8)\nonumber\\
&-&\left(\frac{\sqrt{2}}{v}\right)^2 \frac{m_D^2}{cos^2\beta}
		\label{Td1}
\end{eqnarray}
\begin{eqnarray}
\delta  T_{d_2}&=&\frac{m_W^2}{v^2}\left(2+\frac{1}{c_w^2}\right)+3\lambda_2+2\lambda_3+\lambda_4+(3\lambda_7+\frac{3}{2}\lambda_9)\nonumber\\
&-&\left(\frac{\sqrt{2}}{v}\right)^2 \frac{m_U^2}{sin^2\beta}
	\label{Td2}
\end{eqnarray}
and for the triplet:
\begin{eqnarray}
	\delta T_t&=&4\frac{m_W^2}{v^2}\left(1+\frac{1}{c_w^2}\right)+2(\lambda_6+\lambda_7)+\lambda_8+\lambda_9+8\bar{\lambda}_8+6\bar{\lambda}_9
	\label{Tt}
\end{eqnarray}
\paragraph*{}
\begin{figure*}[hbtp]
	\centering	
	\includegraphics[height =5cm,width=5cm]{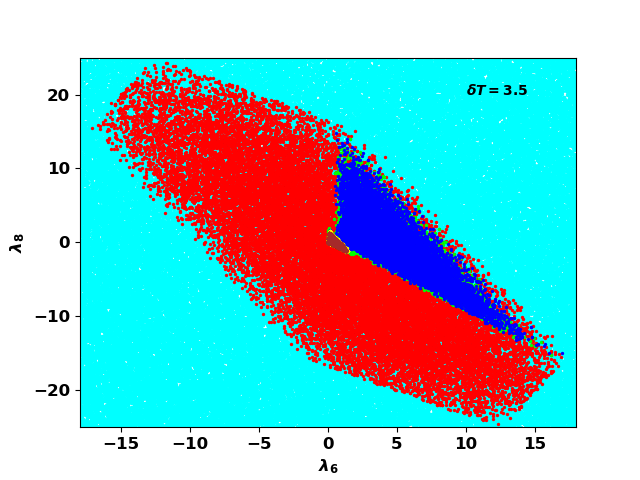}
    \includegraphics[height =5cm,width=5cm]{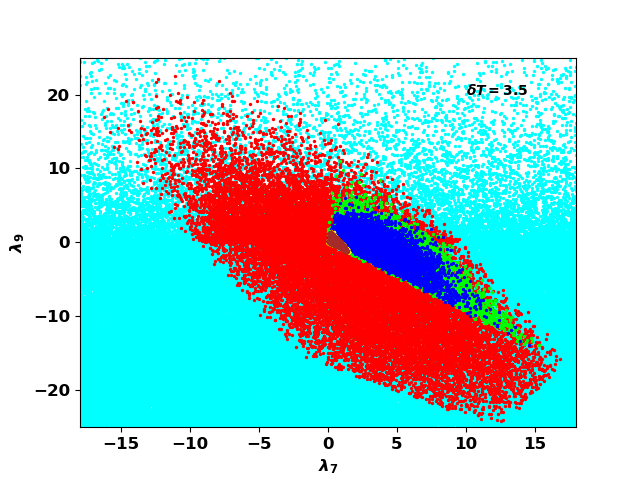}
     \includegraphics[height =5cm,width=5cm]{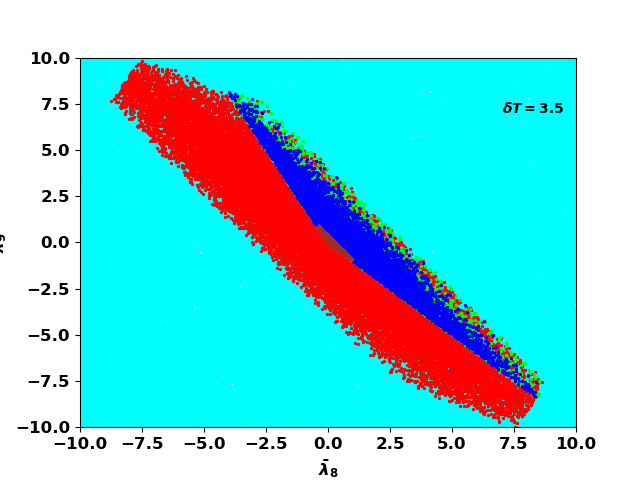} 
	\caption[titre court]{The allowed regions in ($\lambda_6$, $\lambda_8$) (left), ($\lambda_7$, $\lambda_9$) (middle) and ($\bar{\lambda}_8$, $\bar{\lambda}_9$) (right) by imposing $C_1$ (\textcolor{blue}{blue}), combined $C_1$ and $EWPO$ (\textcolor{plum}{plum}), $C_2$ (\textcolor{yellow}{yellow}) and $C_3$ (\textcolor{brown}{brown}) constraints}
	\label{fig:1}
\end{figure*} 
At this point, several comments and remarks are in order:
\begin{itemize}
\item[$\bullet$] The potential parameter $\lambda_1$ is missing in $\delta T_{d_2}$, and similarly $\lambda_2$ is absent in $\delta T_{d_1}$, while $\delta T_t$ excludes both $\lambda_1$ and $\lambda_2$. This can be  explained via Eqs. (\ref{Td1}, \ref{Td2}, \ref{Tt}) which show that $\lambda_1$ and $\lambda_2$ couple solely to $H_1$ and $H_2$ field resepctively. The couplings $\bar{\lambda}_8$ and $\bar{\lambda}_9$, originating from the triplet scalar, rather manifest themselves in $\delta T_t$.  

\item[$\bullet$] The couplings $\lambda_6$ and $\lambda_8$ ($\lambda_7$ and $\lambda_9$) do not appear in $\delta T_{d_2}$ ($\delta T_{d_1}$) since they are only related to $H_1$ ($H_2$) potential terms.

\item[$\bullet$] The Veltman conditions for the Higgs Triplet Model with $Y=2$ reported in \cite{Chabab:2015nel} can be recovered when some of the couplings vanish, and $v_1$ traded for $v_d$ in Eqs.(\ref{Td1}, \ref{Td2}, \ref{Tt}). Similarly, the Veltman conditions in $2HDM$ \cite{Darvishi:2017bhf,Grzadkowski:2009mj,Grzadkowski:2010dn,drozd2012multi,Bazzocchi:2012de,Masina:2013wja,Chakraborty:2014oma,Biswas:2014uba,chowdhury2015global} are reproduced when the couplings that fingerprint the scalar triplet in the Lagrangian, namely $\lambda_6$, $\lambda_7$, $\lambda_8$, $\lambda_9$, $\bar{\lambda}_8$, $\bar{\lambda}_9$, are removed from Eqs. (\ref{Td1}, \ref{Td2}, \ref{Tt}).
\end{itemize}
To proceed with the implementation of the $VC$’s in the parameter space and in the subsequent scans one usually assumes that the deviation $\delta T$ should not exceed the Higgs mass scale  (Veltman's theorem $\delta m_H^{2} /m_H^{2} \ll 1$). To determine the most appropriate values of $\delta T$ for the phenomenological analysis, we first allowed $\delta T$ to lie within the conservative range of $0.1$ to $4$ GeV.  Then we identified the following features:
\begin{itemize}
\item[$\bullet$]  Naturalness constraints are stronger than the other theoretical conditions.
\item[$\bullet$] The deviations $\delta T$ must exceed $2.6$ GeV to maintain a viable model.
\item[$\bullet$] The sample of generated points that satisfy all constraints, including $VC$, is significant when $\delta T$ is about $3.5$ GeV.

 Therefore, to balance between these two requirements (Veltman's theorem and the model viability), we assume thereafter that the deviations in $\delta T$ should not exceed $3.5$ GeV.

\end{itemize}

\section{CONSTRAINTS FROM $B\to X_s\gamma$}
\label{BtoXga}
The  $B\to X_s\gamma$ decay is an important process for constraining new physics, as it is highly sensitive to physics beyond the Standard Model. Therefore, any model that aims to predict a light charged Higgs boson must also be consistent with the experimental constraints on this decay. This feature has been observed in several models with charged Higgs states, such as the two Higgs doublet model extended by a real triplet scalar field \cite{Ait-Ouazghour:2020slc} where a charged Higgs with a mass smaller than the $200$ GeV has been predicted. 
In this section, following  \cite{borzumati1998two}, we investigate whether the $2HDMcT$ model has the potential  to accommodate a light charged Higgs, while being consistent with $BR(\bar{B}\to X_s\gamma)$ decay experimental limits. 

\paragraph*{}
From  Belle \cite{Belle:2014nmp}, Babar \cite{BaBar:2007yhb,BaBar:2012fqh,BaBar:2012eja} and CLEO \cite{CLEO:2001gsa} measurements extrapolated down to $E_0 = 1.6$ GeV, the experimental world average evaluated by HFAG
~\cite{HFLAV:2022pwe} reads:
\begin{eqnarray}
	{\cal BR}(\overline{B}\to X_s\gamma)|_{E_\gamma >1.6\mathrm{~GeV}}=\left(3.40\pm0.17\right)\times 10^{-4}
\end{eqnarray} 

In this context, as a first step towards exploring the viability of the $2HDMcT$,  we first calculate the $BR(\bar{B}\to X_s\gamma)$ decay rate at $NLO$ where the two singly charged Higgs states are both involved. 

\paragraph*{}
One can write the branching ratio of the radiative $B\to X_s\gamma$ decay process at $NLO$ in QCD as:
\begin{eqnarray}
BR(B \rightarrow X_s \gamma) = \frac{\Gamma(B\rightarrow X_s \gamma)}{\Gamma_{SL}}BR_{SL}	
\end{eqnarray}

where $BR_{SL}$ is the measured semileptonic branching ratio, whereas the semileptonic decay width $\Gamma_{SL}$ is expressed by:
\begin{eqnarray}
\Gamma_{SL} = \frac{G_F^2}{192 \pi^3}|V_{cb}|^2m_b^5 g(z)\left(1-\frac{2\alpha_s(\bar\mu_b)}{3\pi}f(z)+\frac{\delta^{NP}_{SL}}{m_b^2}\right);\;\;\;\;z=\frac{m_c^2}{m_b^2}
\end{eqnarray}
The phase space function $g(z)$, the (approximated) QCD–radiation function $f(z)$ and the non–perturbative correction $\delta^{NP}_{SL}$ are defined in \cite{borzumati1998two}.
The decay $\Gamma(B\rightarrow X_s \gamma)$ at $NLO$ is given by:
\begin{eqnarray}
\label{widthrad}
\Gamma(B\rightarrow X_s \gamma)=\frac{G_F^2}{32\pi^4}  
\vert V_{ts}^\star V_{tb}\vert ^2 
\alpha_{em} m_b^5\left( 
|\hd|^2 + A + 
\frac{\dnp_\gamma}{m_b^2} 
\vert C_7^{0,\,{\rm eff}}(\mub) \vert^2\right. \\
& & 
\hspace{-6.5cm}\left.+ 
\frac{\dnp_c}{m_c^2}  
{\rm Re} 
\left(
\left[C_7^{0,\,{\rm eff}}(\mub)\right]^*
\left(C_2^{0,\,{\rm eff}}(\mub)
-\frac{1}{6} C_1^{0,\,{\rm eff}}(\mub)\right)
\right)
\right).\nonumber
\end{eqnarray}
where the amplitudes $A$ and  $\bar D$ read as,
\begin{eqnarray}
A=\frac{\alpha_s(\mu_b)}{\pi}\sum_{i,j=1;i \leq j}^{8}Re\left( C^{0,eff}_i(\mu_b)[C^{0,eff}_j(\mu_b)]^*f_{ij}\right)
\end{eqnarray}
\begin{eqnarray}
\bar D = C^{0,eff}_7(\mu_b) + \frac{\alpha_s(\mu_b)}{4\pi}\left( C^{1,eff}_7+V(\mu_b)\right)
\end{eqnarray}
with $C^{0,eff}_7(\mu_b)$ and $C^{1,eff}_7(\mu_b)$ defined as :
\begin{eqnarray}
C^{0,eff}_7(\mu_b)= \eta^{\frac{16}{23}}C^{0,eff}_7(\mu_w)+\frac{8}{3}(\eta^{\frac{14}{23}}-\eta^{\frac{16}{23}})C^{0,eff}_8(\mu_w)+\sum_{i=1}^{8}h_i\eta^{a_i}
\end{eqnarray}
\begin{eqnarray} 
C^{1,\,{\rm eff}}_7(\mub) & = & 
\eta^{\f{39}{23}} C^{1,\,{\rm eff}}_7(\muw) + 
\f{8}{3} \left( \eta^{\f{37}{23}} - \eta^{\f{39}{23}} \right) 
C^{1,\,{\rm eff}}_8(\muw) \nonumber\\
&+&\left( \f{297664}{14283}   \eta^{\f{16}{23}}
-\f{7164416}{357075} \eta^{\f{14}{23}} 
+\f{256868}{14283}   \eta^{\f{37}{23}}\right) C^{0,\,{\rm eff}}_8(\muw)\nonumber\\
&-&
\f{6698884}{357075} \eta^{\f{39}{23}} C^{0,\,{\rm eff}}_8(\muw)+\, \f{37208}{4761} \left(\eta^{\f{39}{23}} -\eta^{\f{16}{23}} \right) 
C^{0,\,{\rm eff}}_7(\muw)\nonumber\\
&+&
\sum_{i=1}^8  \left(
e_i \,\eta \,C^{1,\,{\rm eff}}_4(\muw)
+ (f_i + k_i \eta) \,C^{0,\,{\rm eff}}_2(\muw)\right) \eta^{a_i}\,, \nonumber\\
&+&
\sum_{i=1}^8 l_i \,\eta\, C^{1,\,{\rm eff}}_1(\muw) \eta^{a_i} \,,
\label{runc7eff1}
\end{eqnarray}

where $V(\mu_b)$, $\eta$, $C^{0,eff}_1(\mu_w)$, $C^{0,eff}_2(\mu_w)$, $C^{1,eff}_1(\mu_w)$, $C^{1,eff}_4(\mu_w)$, and the vectors $a_i$, $h_i$, $e_i$, $f_i$, $k_i$, $l_i$ are reported in \cite{borzumati1998two}.

The effective Wilson coefficient $C^{1,eff}_7(\mu_w)$ at the  $\mu_W$ scale
 is given by:

\begin{eqnarray}
C^{1,eff}_7(\mu_w)=C^{1,eff}_{7,SM}+\sum_{a=1,2} |Y_a|^2C^{1,eff}_{7,Y_aY_a}+X_aY_a^*C^{1,eff}_{7,X_aY_a}
\end{eqnarray}
The coefficient $C_{7,SM}^0(\mu_w)$ is a function of $x=m^2_t/M_w^2$ , while $C^{1,eff}_{7,Y_aY_a}(\mu_w)$ and $C^{1,eff}_{7,X_aY_a}(\mu_w)$ $(a=1,\;2 )$ are functions of $y=m_t^2/m^2_{H^\pm_a}$ ($H^\pm_a=H^\pm_1,\;H^\pm_2$); their explicit forms are summarized in \cite{borzumati1998two}. In our model, the couplings $Y_1$, $Y_2$, $X_1$ and $X_2$ read as:

\begin{eqnarray}
Y_1= \frac{C_{22}}{\sin \beta},\;\; Y_2= \frac{C_{32}}{\sin \beta},\;\;X_1=-\frac{C_{21}}{\cos \beta}\;\; and \;\; 	X_2=-\frac{C_{31}}{\cos \beta}
\end{eqnarray}
where the elements of rotation matrix  $C_{ij}$ can be collected from \cite{Ouazghour:2018mld},
\paragraph*{}
Now that we have defined all the necessary ingredients, we can examine the branching ratio of $B \rightarrow X_s \gamma$ radiative decay  at $NLO$ within $2HDMcT$. 

\begin{figure}[hbtp]
	\centering	
	\includegraphics[height =4.6cm,width=4.28cm]{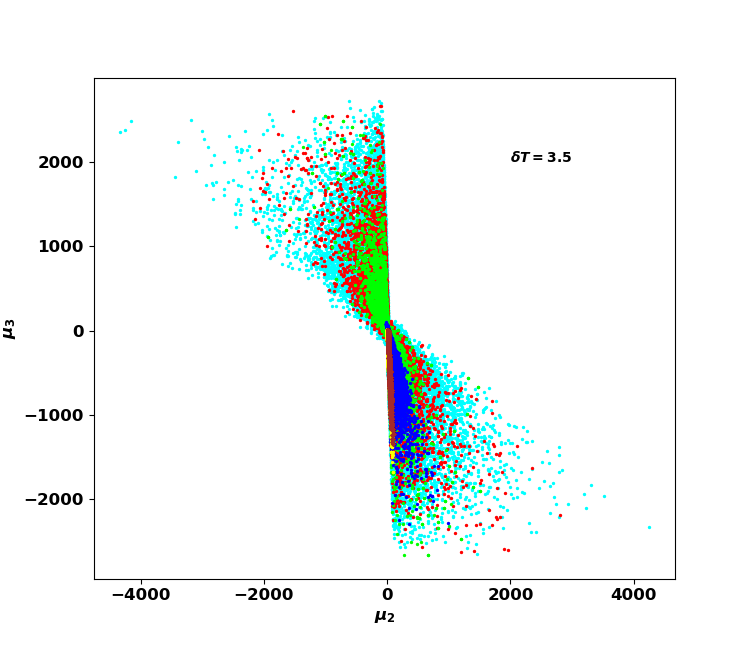}
\includegraphics[height =5cm,width=4.28cm]{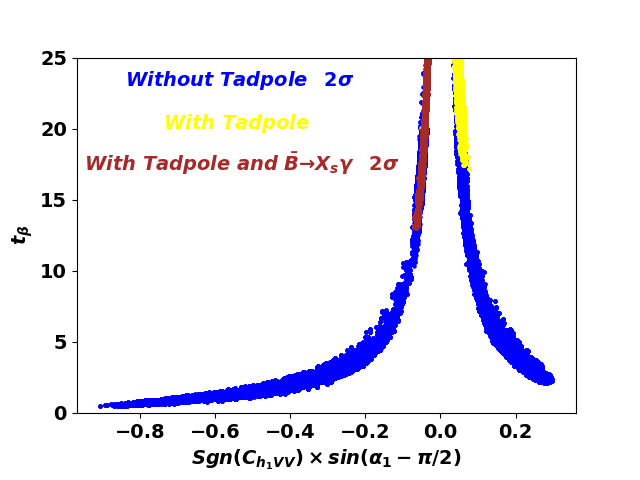}
	\caption[titre court]{The allowed regions in the planes $\mu_2$ vs $\mu_3$ (left) and ($sgn(C_V^{h_1})\sin(\alpha_1-\pi/2)$, $\tan\beta$) (right). All theoretical and experimental constraints are taken into account with color captions similar to Fig~[\ref{fig:1}].}
	\label{fig:2}
\end{figure} 
\begin{figure*}[hbtp]
	\centering	
	\includegraphics[height =5cm,width=5cm]{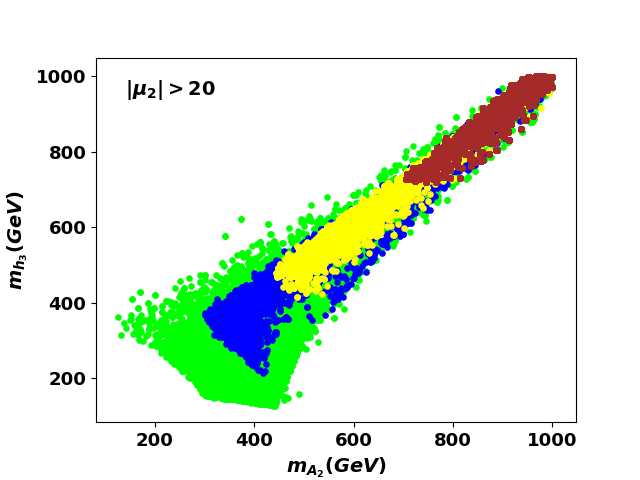}
\includegraphics[height =5cm,width=5cm]{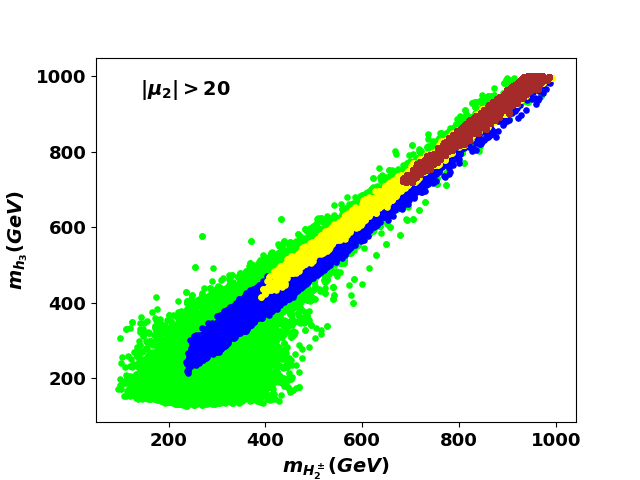}
\includegraphics[height =5cm,width=5cm]{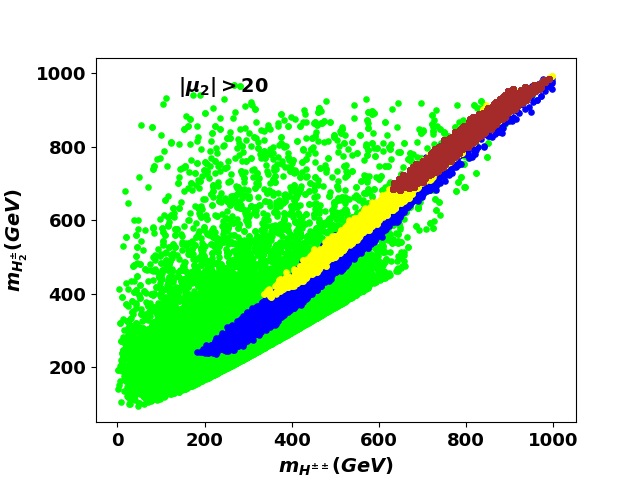}
\includegraphics[height =5cm,width=5cm]{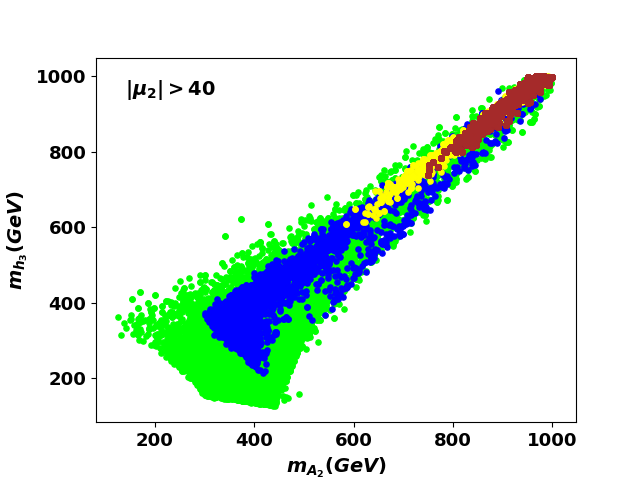}
\includegraphics[height =5cm,width=5cm]{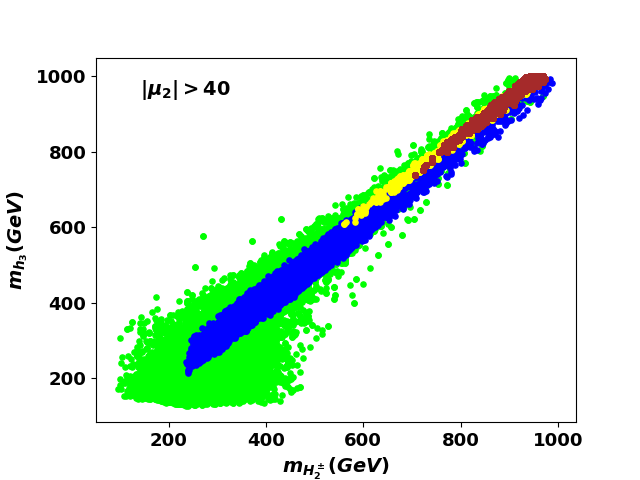}
\includegraphics[height =5cm,width=5cm]{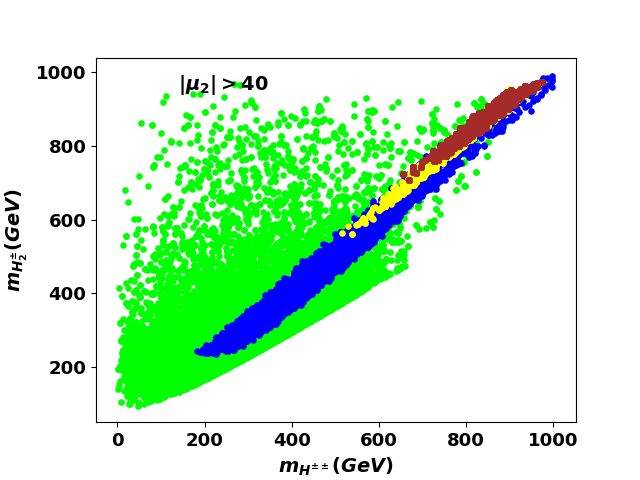}
\includegraphics[height =5cm,width=5cm]{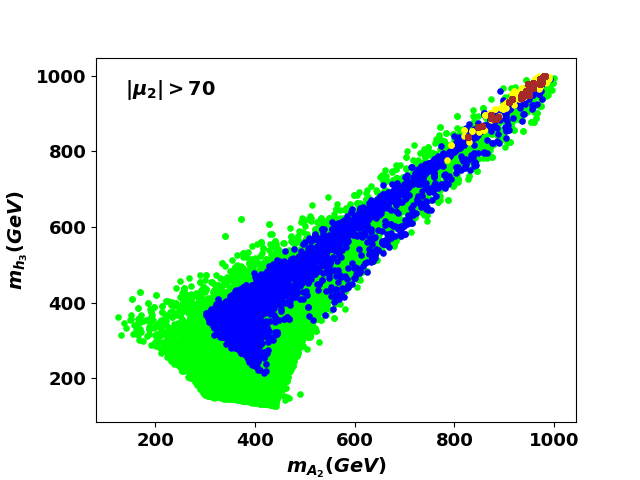}
\includegraphics[height =5cm,width=5cm]{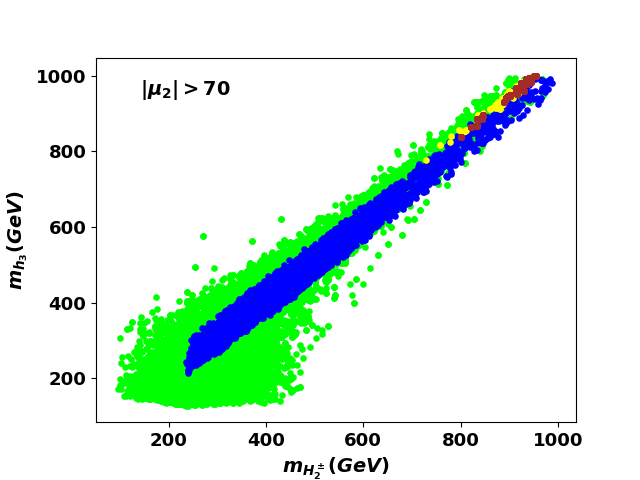}
\includegraphics[height =5cm,width=5cm]{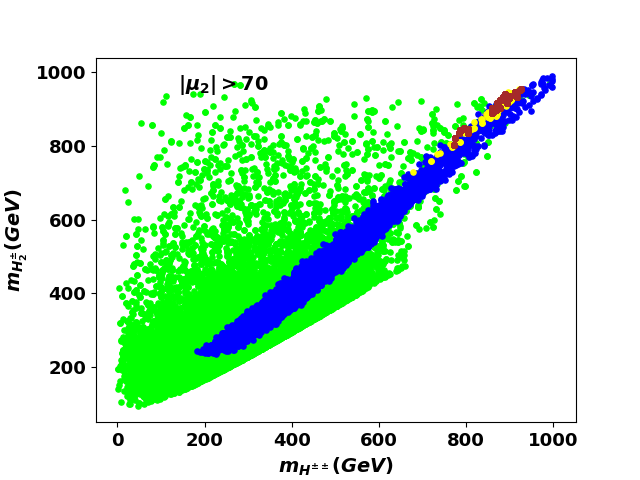}
	\caption{Allowed Higgs mass ranges in the planes $m_{\phi_i}$ vs $m_{\phi_j}$  ($\phi_i=$ $h_3$, $A_2$, $\phi_i=H^\pm_2$ and $\phi_i=H^{\pm\pm}$). All theoretical and experimental constraints are taken into account with color captions similar to Fig.~[\ref{fig:1}].}
	\label{fig:4}
\end{figure*}
In Fig.~[\ref{fig:constraints}], we plot the branching ratio of $B \rightarrow X_s \gamma$ in the $2HDMcT$ as a function of $m_{H^\pm_1}$. $\tan \beta$ is respectively fixed  to $1$ (left panel), $2$ (center panel) and $15$ (right panel) with the mass difference $\Delta  M_{H^\pm}=m_{H^\pm_2}-m_{H^\pm_1}=20,\;50,\;and \;100$ GeV. We also used different values of the mixing angle  $\theta_3^\pm$ ranging from $0.05$ corresponding to a mostly triplet $H_2^\pm  (H_1^\pm )$, to $ \pi/4$ for a nearly equal contributions from doublets and triplet scalars to the charged Higgs bosons. For a comparative analysis, $2HDM$ prediction is also illustrated in black color. First, we see that the difference between the predictions in $2HDM$ and $2HDMcT$ becomes slightly larger as the mass difference  $\Delta M_{H^\pm}$  or the $\theta_3^\pm$ increase. Besides, Fig.~[\ref{fig:constraints}] uncovers two main features: 
 \begin{itemize}
 	\item The Higgs mass range $H_1^\pm$ satisfying the $B \rightarrow X_s \gamma$ constraint is almost insensitive to $\tan \beta$.  
	\item The excluded range of the $m_{H^\pm_1}$ by $B \rightarrow X_s \gamma$ reduces when $\Delta  M_{H^\pm}$ gets large.
\end{itemize}	

\section{ANALYSIS: IMPLICATIONS FOR THE PHYSICAL SCALAR MASSES}
\label{result}
\paragraph*{}
In this section, we focus on identifying viable regions of parameter space that satisfy all previously derived theoretical and experimental constraints. We pay special attention to the impact of three Veltman conditions $VC$'s as the experimental constraints arising from the $B\to X_s\gamma$ decay.  Throughout the paper, $h_1$ is identified as the SM Higgs-like particle observed at the LHC with $m_{h_1}=125\;GeV$. \\
The input parameters, subject to the various constraints mentioned above, which we will use in the subsequent analysis, are summarized as follows:
	\begin{eqnarray}
	\begin{matrix}
		m_{h_1}=125\,\,\text{GeV},\,\,m_{h_1}\leq m_{h_2}\leq m_{h_3}\leq 1\,\text{TeV},\,\,\\
		80\,\text{GeV}\leq m_{H^{\pm\pm}}\leq 1\,\text{TeV},\,\, \frac{-\pi}{2}\leq \alpha_{1}\leq \frac{\pi}{2},\,\, -0.1\leq \alpha_{2,3}\leq 0.1\\
		0.5\leq \tan\beta\leq 30,\,\,-1\leq \mu_1\leq 1,\,\,0\leq v_t\leq 1 \,\text{GeV},\,\,\,\lambda_1\approx0.15\\,\,\,\lambda_3\approx1.6,\,\,\lambda_4\approx 1.6,\,\,-8\pi\leq \lambda_{6,7}\leq 8\pi,\,\,-8\pi\leq \lambda_8\leq 8\pi,\,\,-8\pi\leq \lambda_9\leq 8\pi
	\end{matrix}
	\end{eqnarray}
Fo commodity purpose, we also classify the theoretical and experimental constraints into three sets :  
\begin{itemize}
	\item[$\bullet$] First set, $C_1$, includes the unitarity, perturbativity, vacuum stability and experimental Higgs boson exclusion limits .
	
	\item[$\bullet$] $C_2$, contains all constraints in $C_1$ in addition to the electroweak precision observables and the modified Veltman conditions.
	\item[$\bullet$] $C_3$ includes  $C_2$ augmented by $B\to X_s\gamma$ constraint at 95$\%$ C.L.
\end{itemize}
\paragraph*{}
We present in Fig.~[\ref{fig:1}] the allowed regions in $\left(\lambda_6\,;\,\lambda_8\right)$, $\left(\lambda_7\,;\,\lambda_9\right)$ and $\left(\bar{\lambda}_8\,;\,\bar{\lambda}_9\right)$ excluded  by various theoretical constraints and experimental measurements. We can see that the allowed regions of these parameter spaces undergo drastic reductions as more constraints are imposed. More specifically, they are sizably shrunk to limited yellow areas when naturalness is invoked. As a result, the potential parameters are allowed to vary within reduced intervals as shown in Table \ref{region_of_parameters}.
\begin{table}[hptb]  
\begin{center}
	\caption{The allowed regions of potential parameters from $C_2$ constraints.}
	\renewcommand{\arraystretch}{1.1}
	\begin{tabular}{|c||c|}
		\hline\hline
		parameters & intervals\\ \hline 
		$\lambda_6$ & $[0,\;1.46]$ \\ \hline 
		$\lambda_7$ & $[0,\;1.49]$\\ \hline
		$\lambda_8$ & $[-1.19,\;1.57]$ \\ \hline
		$\lambda_9$ & $[-1.34,\;1.57]$  \\ \hline
		$\bar{\lambda}_8$ & $[-0.41,\;0.96]$  \\ \hline
		$\bar{\lambda}_9$ & $[-0.9,\;0.86]$  \\ \hline
		\hline
	\end{tabular}
	\label{region_of_parameters}
\end{center}
\end{table}
Also we can observe that the parameters $\lambda_i$ ($i=6,7,8\,\,and\,\,9$) and $\bar{\lambda}_{8,9}$ are fully insensitive to the constraints from measurements $\bar{B}\to X_s\gamma$ decay rate. 
\paragraph*{}
As for the $\mu_2$ and $\mu_3$ parameters plotted in the left panel of  Fig.~[\ref{fig:2}], we see that they are more responsive to the naturalness induced conditions, particularly $T_t$, than to the other theoretical constraints.  As a result, our analysis shows that the allowed region of $\mu_2$ parameter space undergoes a drastic reduction $\mu_2 \in [-83,\;82]$,
whereas $\mu_3$ lies within the allowed interval  $\mu_3 \in [-1230,\;1698]$. The right panel of Fig.~[\ref{fig:2}] illustrates the scatter plot in $\tan\beta$ and $sgn(C_V^{h_1})\sin(\alpha_1-\pi/2)$ for $\Delta \chi^2 < 5.99$. When naturalness is switched off, the corresponding generated samples are shown in blue at $2\sigma$ whereas the yellow dots signal inclusion of the Veltman conditions. This graph indicates that only $\tan \beta > 13$ and $\alpha_1 \in  [-1.57 ; -1.51]U[1.51 ; 1.57]$  are compatible with all constraints. At this stage, it is worth noticing that the left branch with $\sin(\alpha_1-\frac{\pi}{2}) < 0$ corresponds to the SM-alignment limit, where the couplings of CP even scalars to gauge bosons are assumed to mimic the SM Higgs coupling. The right branch with  $\sin(\alpha_1-\frac{\pi}{2}) > 0$ represents the wrong sign Yukaya coupling limit. If, additionally, constraints from the measurements of the $\bar{B}\to X_s \gamma$ decay rate at $95\%$ C.L. intervenes, then only $\alpha_1$ positive survives. 
\paragraph*{}
Fig.~[\ref{fig:4}] displays the allowed region in the $(m_{A_2}\;,\;m_{h_3})$, $(m_{H^\pm_2}\;,\;m_{h_3})$ and $(m_{H^{\pm\pm}}\;,\;m_{H^{\pm}_2})$ for three ranges of $|\mu_2| > 20,$ $40$ and $70$. Firstly, we notice a significant correlation between the mass of the Higgs boson $h_3$ and the mass of the Higgs boson $A_2$. This correlation arises from the assumptions we utilized to calculate the oblique parameters. Secondly, we find that the relevant parameter regions obeying $C_2$ constraints are confined within the yellow areas. The latter extents are proportional to the choice of $\mu_2$ range: When $\mu_2$ values are in the vicinity of zero, the masses are almost insensitive to the naturalness (set $C_3$). However,  if $\mu_2$ moves away from zero, the mass ranges are clearly reduced by $VC$ conditions. The  brown color represents the surviving parameter regions to all constraints. Therefore as summarized by Table \ref{table:masses}, we can conclude that the naturalness has a substantial impact on both the lower bounds of the heavy Higgs spectra and on the upper bounds of the light Higgs bosons masses.  It is worth noticing that this general trend persists even if the mass scale, initially fixed to $1$ TeV, is shifted up.  More specifically, we see that the lower bounds of the Higgs mass ranges  are insensitive to the increase in scale, while the upper bounds are evidently pushed up to higher values, as expected.
\begin{table}\centering
	\caption{Higgs masses allowed intervals from various constraints, including the modified Veltman conditions for  $\delta T=3.5$ GeV}
	\centering 
	\begin{tabular}{l|c|c|c}
		\cline{1-4}
		\\
		\toprule
		\textbf{$m_\phi$} & \textbf{Unitarity} & \textbf{Unitarity+BFB} & \textbf{$\;C_1$ constraints}\\
		\midrule
		\cline{1-4}
		\\
		$m_{h_2}$&$[126\,;\,990]$&$[126\,;\,910]$&$[137\,;\,905]$\\
		$m_{h_3}$&$[126\,;\,1000]$&$[127\,;\,1000]$&$[266\,;\,999]$\\
		$m_{A_1}$&$[80\,;\,992]$&[80\,;\,930]&$[128\,;\,903]$\\
		$m_{A_2}$&$[126\,;\,1000]$&$[127\,;\,1000]$&$[266\,;\,999]$\\
		$m_{H^\pm_1}$&$[80\,;\,991]$&$[80\,;\,933]$&	$[218\,;\,913]$\\
		$m_{H^\pm_2}$&$[88\,;\,998]$&$[95\,;\,991]$&	$[276\,;\,982]$\\
		$m_{H^{\pm\pm}}$&$[80\,;\,1000]$&$[80\,;\,1000]$&$[235\,;\,997]$\\
		\bottomrule
	\end{tabular}
	\begin{tabular}{l|c|c|c}
		\cline{1-4}
		\\
		\toprule
		\textbf{$m_\phi$} & \multicolumn{2}{|c}{\textbf{$\,\,\,\;\,\,\;\,\,\;\,\,\;\,\,\;\;\;\;\;\;\;\;\;\;\;$$C_2$ constraints}} \\ \cline{2-4}
		& $|\mu_2|>20$ \textbf{GeV}& $|\mu_2|>40$ \textbf{GeV}& $|\mu_2|>70$ \textbf{GeV}\\
		\midrule
		\cline{1-4}
		\\
		$m_{h_2}$&$[145\,;\,930]$& $[160\,;\,880]$&$[167\,;\,795]$\\
		$m_{h_3}$& $[415\,;\,999]$& $[607\,;\,999]$&$[778\,;\,999]$\\
		$m_{A_1}$&$[145\,;\,950]$& $[170\,;\,920]$&$[236\,;\,840]$\\
		$m_{A_2}$&$[415\,;\,999]$& $[607\,;\,999]$&$[778\,;\,999]$\\
		$m_{H^\pm_1}$&$[222\,;\,980]$& $[266\,;\,918]$&$[328\,;\,840]$\\
		$m_{H^\pm_2}$& $[391\,;\,987]$& $[560\,;\,976]$&$[728\,;\,956]$\\
		$m_{H^{\pm\pm}}$&$[335\,;\,995]$& $[515\,;\,976]$&$[679\,;\,928]$\\
		\bottomrule
	\end{tabular}
\begin{tabular}{l|c|c|c}
	\cline{1-4}
	\\
	\toprule
	\textbf{$m_\phi$} & \multicolumn{2}{|c}{$\,\,\,\;\,\,\;\,\,\;\,\,\;\,\;\;\;\;\;\;\;\;\;\;\;$\textbf{$C_3$ constraints}} \\ \cline{2-4}
	& $|\mu_2|>20$ \textbf{GeV}& $|\mu_2|>40$ \textbf{GeV} & $|\mu_2|>70$ \textbf{GeV}\\
	\midrule
	\cline{1-4}
	\\
	$m_{h_2}$&$[487\,;\,927]$& $[490\,;\,879]$&$[511\,;\,790]$\\
	$m_{h_3}$& $[719\,;\,999]$& $[737\,;\,999]$&$[837\,;\,982]$\\
	$m_{A_1}$&$[576\,;\,947]$& $[576\,;\,910]$&$[596\,;\,838]$\\
	$m_{A_2}$&$[719\,;\,999]$& $[737\,;\,999]$&$[837\,;\,982]$\\
	$m_{H^\pm_1}$&$[592\,;\,946]$& $[599\,;\,915]$&$[623\,;\,839]$\\
	$m_{H^\pm_2}$& $[683\,;\,985]$& $[706\,;\,973]$&$[801\,;\,954]$\\
	$m_{H^{\pm\pm}}$&$[632\,;\,990]$& $[654\,;\,973]$&$[772\,;\,925]$\\
	\bottomrule
\end{tabular}
	\label{table:masses}
\end{table}
\section{CONCLUSION}
\label{conlusion}
In this paper, we have extended the work \cite{Ouazghour:2018mld} by investigating naturalness problem and B physics constraints within the context of two Higgs doublets type II seesaw model  ($2HDMcT$). We have derived the modified Vetman conditions in $2HDMcT$ and showed that the naturalness problem might be avoided  within the model parameter space at the electroweak scale. Next, we have calculated the branching ratio of $\bar{B}\to X_s\gamma$  radiative decay at the $NLO$ and showed that  $\bar{B}\to X_s\gamma$ constraint can only be accommodated provided that the parameter $\alpha_1$ is always positive and  the Higgs mass hierarchy $m_{h_{3}}>m_{H_2^\pm}>m_{H^{\pm\pm}}$ satisfied. Th main outcome of our current analysis is to show how these additional new constraints,  $VC$ and $\bar{B}\to X_s\gamma$,  induce significant delimitations of the $2HDMcT$  parameter space with respect to the results reported in \cite{Ouazghour:2018mld}.   As a consequence, it is demonstrated that the Higgs spectrum is utterly reshaped with  the heavy Higgs $h_3$, $A_2$, $H_2^\pm$ and $H^{\pm\pm}$ masses significantly affected. Three benchmark scenarios corresponding to $\mu_2 > 20, 40, 70$ GeV have been explored. For $\mu_2 > 40$ GeV, we found that the lower mass limits of nonstandard Higgs  bosons ($h_2$, $h_3$, $A_1$, $A_2$, $H_1^\pm$, $H_2^\pm$, $H^{\pm\pm}$) are pushed up to higher values ranging from $490$ to $750$ GeV.

\newpage
\pagebreak
\bibliographystyle{JHEP}
\bibliography{references}
\end{document}